\documentstyle[buckow]{article}
\let\du=\du                     

\def\a{\alpha}

\def\d{\delta}

\def\l{\lambda}

\def\p{\pi}

\def\x{\xi}

\def\ve{\varepsilon}

\def\cd{{\cal D}}

\def\bo{{\raise-.3ex\hbox{\large$\Box$}}}               
\def\pa{\partial}                                       
\def\TH{{\raise.2ex\hbox{$\displaystyle \bigodot$}\mskip-4.7mu \llap H \;}}
\def\face{{\raise.2ex\hbox{$\displaystyle \bigodot$}\mskip-2.2mu \llap {$\ddot
        \smile$}}}                                      

\def\VEV#1{\left\langle #1\right\rangle}        
\def\abs#1{\left| #1\right|}                    
\def\leftrightarrowfill{$\mathsurround=0pt \mathord\leftarrow \mkern-6mu
        \cleaders\hbox{$\mkern-2mu \mathord- \mkern-2mu$}\hfill
        \mkern-6mu \mathord\rightarrow$}
\def\dvec#1{\vbox{\ialign{##\crcr
        \leftrightarrowfill\crcr\noalign{\kern-1pt\nointerlineskip}
        $\hfil\displaystyle{#1}\hfil$\crcr}}}           

\def\frac#1#2{{\textstyle{#1\over\vphantom2\smash{\raise.20ex
        \hbox{$\scriptstyle{#2}$}}}}}                   
\def\sfrac#1#2{{\vphantom1\smash{\lower.5ex\hbox{\small$#1$}}\over
        \vphantom1\smash{\raise.4ex\hbox{\small$#2$}}}} 
\def\bfrac#1#2{{\vphantom1\smash{\lower.5ex\hbox{$#1$}}\over
        \vphantom1\smash{\raise.3ex\hbox{$#2$}}}}       
\def\afrac#1#2{{\vphantom1\smash{\lower.5ex\hbox{$#1$}}\over#2}}    

\def\[{\lfloor{\hskip 0.35pt}\!\!\!\lceil}
\def\]{\rfloor{\hskip 0.35pt}\!\!\!\rceil}

\def\du#1#2{_{#1}{}^{#2}}

\def\Tr{{\rm Tr}}

\def\fracmm#1#2{{{#1}\over{#2}}}

\def\low#1{{\raise -3pt\hbox{${\hskip 0.75pt}\!_{#1}$}}}

\newskip\humongous \humongous=0pt plus 1000pt minus 1000pt

\newif\ifdtup

\def\pl#1#2#3{Phys.~Lett.~{\bf {#1}B} (19{#2}) #3}
\def\np#1#2#3{Nucl.~Phys.~{\bf B{#1}} (19{#2}) #3}

\def\ibid#1#2#3{{\it ibid.}~{\bf {#1}} (19{#2}) #3}

\begin {document}
\def\titleline{
AdS/CFT correspondence and coincident D-6-branes
}
\def\authors{
Sergei V. Ketov
}
\def\addresses{
Institut f\"ur Theoretische Physik, Universit\"at Hannover\\
Appelstra\ss{}e 2, 30167 Hannover, Germany
}
\def\abstracttext{
A relation between confinement and Maldacena conjecture is briefly
discussed. The gauge symmetry enhancement for two coincident D-6-branes
is analyzed from the viewpoint of the hypermultiplet low-energy 
effective action given by the N=2 supersymmetric non-linear sigma-model
with the Eguchi-Hanson (ALE) target space.
}
\large
\makefront

The most attractive mechanism of color confinement in QCD is known to be
the dual (type II) superconductivity, i.e. a creation of color-electric
fluxes (or strings) having quarks at their ends \cite{hooft,mandel}.
The usual type-II superconductivity (i.e. the confinement of magnetic charges)
is known to be a solution to the standard Landau-Ginzburg theory, 
whereas the QCD confinement is supposed to be a non-perturbative solution 
to a (1+3)-dimensional quantum $SU(N_c)$ gauge field theory with $N_c=3$. 

The formal proof of the confinement in QCD amounts to a derivation of the area
law for a Wilson loop $W[C]$. It may be based on the following `string' Ansatz 
\cite{polbook}:
$$ W[C] \sim \int_{{\rm surfaces~}B,\atop \pa B=C}\; 
\exp\left(-S_{\rm string}\right)~~.\eqno(1)$$
Eq.~(1) clearly shows that the effective degrees of freedom (or collective
coordinates) in QCD at strong coupling (in the infra-red) are strings whose 
worldsheets are given by surfaces $B$ and whose dynamics is governed by (still 
unknown) action $S_{\rm string}$. The fundamental (Schwinger-Dyson) equations 
of QCD can be reformulated into the equivalent infinite chain of equations for 
the Wilson loops \cite{mig}. The chain of loop equations drastically simplifies
 at large number of colors $N_c$ (i.e. when only planar Feynman 
graphs are taken into account) to a {\it single} closed equation known as the 
{\it Makeenko-Migdal} loop equation \cite{mm}. Eq.~(1) is, therefore, just the 
Ansatz for a solution to the MM loop equation in terms of some string action 
$S_{\rm string}$ to be determined out of it. The main (yet unsolved) problems 
in a realization of this program in the past were (i) taking into account 
quantum renormalization in the MM equation, and (ii) determining the 
corresponding string action \cite{mpriv}. The first problem does not arise if
one replaces QCD by the N=4 supersymmetric Yang-Mills (SYM) theory and 
considers the N=4 supersymmetric MM-type loop equation instead of the original
(N=0) one, just because of the well-known fact that, being a scale invariant 
quantum field theory in 1+3 spacetime dimensions, the N=4 SYM does not 
renormalize at all. The recent Maldacena conjecture \cite{mal}, claiming that
the N=4 SYM theory is dual to the IIB superstring theory in the 
$AdS_5\times S^5$ background, can then be interpreted as the particular Ansatz
$S_{\rm string}=S_{IIB/AdS_5\times S^5}$ for a solution to the N=4 super-MM 
loop equation in the form of eq.~(1). Within the Maldacena conjecture, the 
(1+3)-dimensional spacetime is identified with the boundary of the 
anti-de-Sitter space $AdS_5$, where
$$ AdS_5=\fracmm{SO(4,2)}{SO(4,1)} \quad {\rm and} \quad 
S^5=\fracmm{SO(6)}{SO(5)}~,\eqno(2)$$
while the coupling constants are related to the $AdS_5$ radius as follows
\cite{mal}:
$$ (\a')^{-2}R^4_{\rm AdS}\sim g^2_{\rm YM}N_c~,
\quad g_{\rm string}\sim g^2_{\rm YM}~.\eqno(3)$$
The proposed duality is a strong-weak coupling duality: 
\begin{itemize}
\item for small $\l=g^2_{\rm YM}N_c$ a perturbative SYM description applies,
\item for large $\l$ a perturbative IIB string/AdS supergravity description 
applies.
\end{itemize}
It is in agreement with the holographic proposal \cite{hsu} since physics in 
the $AdS_5$ bulk is supposed to be encoded in terms of the field theory defined
on the $AdS_5$ boundary. The quantum N=4 SYM theory is conformally invariant, 
while its rigid symmetry is given by the supergroup $SU(2,2|4)$ 
that contains 32
supercharges. The isometries of $AdS_5\times S^5$ form the group $SO(4,2)\times
SO(6)\cong SU(2,2)\times SU(4)$ whose extension in the AdS supergravity is also
given by  $SU(2,2|4)$. In addition, both the N=4 SYM and type-IIB superstrings
are believed to be self-dual under the S-duality group $SL(2,{\bf Z})$. In more
practical terms, this CFT/AdS correspondence is just a one-to-one 
correspondence \cite{gkp} between the N=4 SYM correlators and the correlators 
of the certain string theory whose action $S_{\rm string}$ is known and whose
correlators can be computed, in principle, by the methods of two-dimensional
conformal field theory \cite{mybook}. Quantum corrections in powers of 
$(\a'\times{\rm curvature})$ on the string theory side
correspond to corrections in powers of $(g^2_{\rm YM}N_c)^{-1/2}$ on the gauge
field theory side, while the string loop corrections are suppressed by
powers of $N_c^{-2}$.

The close connection between IIB strings and N=4 SYM is also known to exist 
within the modern brane technology. The type-IIB supergravity (= the low-energy
effective field theory of IIB strings) admits extended solitonic BPS-like 
classical solutions known as D-3-branes and D-strings (or `mesons') \cite{host}.
These solutions can spontaneously break the conformal invariance in the 
non-perturbative N=4 SYM theory, and thus may be useful for a simulation of
confinement. When $N_c$ parallel and similarly oriented
D-3-branes coincide, the low-energy effective field theory action in their
common worldvolume appears to be the N=4 SYM with the gauge group $U(N_c)$
\cite{witten}. The brane picture thus provides a classical resolution to the
non-perturbative N=4 SYM, in the very similar way as the M-theory 5-brane
classical dynamics yields exact Seiberg-Witten-type solutions to N=2 
supersymmetric quantum gauge field theories (see e.g., ref.~\cite{myrev} for a
review). The 11-dimensional M-theory compactified on a 2-dimensional torus is 
known to be dual to the 10-dimensional type-IIB strings compactified on a 
circle \cite{schw}, so that the relevant phenomenon is just the gauge symmetry
enhancement for coincident D-branes alone. We would like to understand it from
the field-theoretical point of view, and distinguish between those elements of
this phenomenon that are of perturbative origin and those elements that are 
truly non-perturbative.

As an example, consider the case of two nearly coincident KK monopoles in M 
theory. The metric is essentially given by the 2-centre Taub-NUT metric (or,
equivalently, the mixed Taub-NUT-Eguchi-Hanson metric) characterized by the
harmonic potential \cite{town} 
$$ H(\vec{y})=\l +\fracmm{1}{2}\left\{ \fracmm{1}{\abs{\vec{y}-\x\vec{e}}}+
\fracmm{1}{\abs{\vec{y}+\x\vec{e}}} \right\}~.\eqno(4)$$
In the limit $\x\to 0$ the homology 2-sphere connecting two KK
monopoles contracts to a point. Since the energy of an M-2-brane wrapped about
this 2-sphere is proportional to its area (of order $\x$), a massless vector
particle (= the zero-mode of the M-2-brane) appears. In the type-IIA picture, 
the ground states of the 6--6 strings connecting two D-6-branes become 
massless if the branes coincide. The net effect is called the non-abelian 
gauge symmetry enhancement: $U(1)\times U(1)\to U(2)$, and it is truly 
non-perturbative \cite{ov}. The gauge fields (and their supersymmetric 
partners) related to non-diagonal gauge symmetry generators are the ground 
states of strings connecting {\it different} D-6-branes, with the masses being
proportional to the distance between the D-branes, whereas those related to 
the Cartan subalgebra generators appear as the massless ground states of the 
strings ending on {\it the same} D-6-brane. The existence of the massless 
ground states associated with Cartan subalgebra can be understood  
perturbatively, as a result of dynamical generation of massless vector 
supermultiplets in the (one-loop) quantum perturbation theory \cite{ket}.

The effective field theory in the D-6-brane worldvolume dimensionally reduced
to four dimensions includes the hyper-K\"ahler non-linear sigma-model (NLSM)
for a self-interacting hypermultiplet, whose NLSM metric is dictated by the
potential (4). This NLSM can be written down in harmonic superspace, in terms 
of two hypermultiplets $q^{A+}$, $A=1,2$, and the auxiliary N=2 vector 
superfield $V^{++}$ as a Lagrange multiplier, 
$$ S_{\rm mixed}[q^A,V^{++}] =\int_{\rm analytic}\left\{
\bar{q}^{A+}D^{++}_Zq^+_A +V^{++}\left( \frac{1}{2}\ve^{AB}\bar{q}^+_Aq^+_B 
+\x^{++} \right) +\frac{1}{4}\l(\bar{q}^{A+}q_A^+)^2\right\}~,\eqno(5)$$
where the Fayet-Iliopoulos term $(\sim \x V)$ has been introduced, while both
hypermultiplets are supposed to have {\it different} masses, $m_1$ and $m_2$,
given by the N=2 central charge $(Z)$ eigenvalues. Eq.~(5) is the gauged NLSM 
over the non-compact coset space $SU(1,1)/U(1)$ parametrized by two 
hypermultiplets. Near the core of D-6-branes the parameter $\l$ becomes 
irrelevant, so that the NLSM target space looks like an ALE space with the 
Eguchi-Hanson metric. Setting $\l=0$ in eq.~(5) results in a formally 
renormalizable four-dimensional `linear' NLSM, that allows us to integrate 
over the hypermultiplets in eq.~(5). All one needs is the N=2 superfield 
hypermultiplet propagator in harmonic superspace \cite{ikz},
$$ i\VEV{q^+(1)\bar{q}^+(2)}=-\fracmm{1}{\bo_1^Z}
(D_1^+)^4(D_2^+)^4\left\{ \d^{12}(Z_1-Z_2)\fracmm{e^{v_Z(2)-v_Z(1)}}{
(u^+_1u^+_2)^3}\right\}~,\eqno(6)$$
where $\bo^Z$ is the Klein-Gordon operator, and $v_Z$ is the co-called `bridge'
(see ref.~\cite{ikz} for details). It is now straightforward to calculate the
one-loop gauge effective action $i\Tr\log(\cd_{Z,V}^{++})$ in the low-energy
approximation \cite{ket}. We find that the N=2 vector supermultiplet, 
introduced in the classical action (5) as the Lagrange multiplier (without 
a kinetic term), becomes dynamical in quantum theory due to a dynamical
generation of its kinetic term
$$ S_{\rm induced}[V^{++}]=-\fracmm{1}{2e^2_{\rm ind}(p)}\int_{\rm chiral}\,
W^2 +{\rm h.c.}~,\eqno(7)$$
where we have introduced the standard N=2 superfield strength 
$W=\int du(\bar{D}^-)^2V^{++}$. The induced (momentum-dependent) gauge coupling
constant appears to be \cite{ket}
$$ \fracmm{1}{e^2_{\rm ind}(p)}=\fracmm{1}{16\p^2}\int^1_0dx\,
\ln \fracmm{m_2^2+p^2x(1-x)}{m_1^2+p^2x(1-x)}= \fracmm{1}{e^2_0}+O(p^2/m^2)~.
\eqno(8)$$
The N=2 central charge providing masses to the hypermultiplets can, therefore,
be understood as the origin of the induced gauge coupling (8).

A generalization to higher unitary groups is straightforward \cite{ket}. The
Hooft limit of large $\l$ appears to be equivalent to $\abs{\x/Z^2}\to 0$ in
our approach. The orthogonal gauge groups may also be considered by introducing
orientifolds and the Atiyah-Hitchin metric in M-theory.

\end{document}